\documentclass{article}

\usepackage{arxiv}

\usepackage[utf8]{inputenc} 
\usepackage[T1]{fontenc}    
\usepackage{hyperref}       
\usepackage{url}            
\usepackage{booktabs}       
\usepackage{amsfonts}       
\usepackage{nicefrac}       
\usepackage{microtype}      
\usepackage{lipsum}		
\usepackage{graphicx}
\usepackage{natbib}
\usepackage{doi}

\title{Relaxation of particle laden interfaces: geometric and preparation effects}

\author{ \href{https://orcid.org/0000-0002-3974-5742}{\includegraphics[scale=0.06]{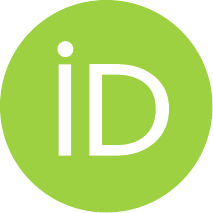}\hspace{1mm}Carole Planchette} \\
	Graz University of Technology\\
	Inffeldgasse 25/F\\
	8010 Graz, Austria \\
	\texttt{carole.planchette@tugraz.at} \\
	\And
	\href{https://orcid.org/0000-0002-0529-2773}{\includegraphics[scale=0.06]{orcid.pdf}\hspace{1mm}Gregor Plohl}\thanks{current address: Infinite Biotech, Zagrebska cesta 20, 2000 Maribor, Slovenia.}  \\
	Graz University of Technology\\
	Inffeldgasse 25/F\\
	8010 Graz, Austria \\
	}

\renewcommand{\shorttitle}{\textit{arXiv} Template}

\hypersetup{
pdftitle={},
pdfsubject={q-bio.NC, q-bio.QM},
pdfauthor={David S.~Hippocampus, Elias D.~Striatum},
pdfkeywords={First keyword, Second keyword, More},
}

\begin{document}
\maketitle

\begin{abstract}
	The relaxation of uniaxially compressed particle rafts through a finite opening on the opposite side is experimentally studied. Three main behaviors are identified. The lowest degree of relaxation corresponds to local unjamming. The other extreme, full relaxation, is characterized by the unjamming of the entire raft. Intermediate relaxation occurs between these extremes. The unjammed domain initially grows along the compression direction with an almost constant width and may extend through the entire raft length. If this occurs, a second phase can begin, during which erosion allows the unjammed channel to develop perpendicular to the compression direction. By using different raft geometries—varied lengths, compression levels, and openings of various widths—we rationalize the occurrence of these behaviors, attributing them to the mechanical robustness of the force chain network. The thresholds for channel formation and erosion are interpreted as ruptures due to excessive shear and elongation, respectively. Comparing results from rafts prepared with three different mixing degrees, we demonstrate that these thresholds are strongly affected by the raft's history and quantify these effects as shifts in the rupture limits.
\end{abstract}


\section{Introduction}
Particle-laden interfaces have raised interest both in the field of material science, where they have long been  used  to stabilize interfaces between two fluids \cite{Ramsden1903, Pickering1907, B606965G, Herzig_2007, Madivala_2009, Rio_2011}, and in the field of physics, where they serve as models for soft glassy materials \cite{sollich1997rheology, hebraud1998mode}. Similarly to insoluble surfactants, they build interfacial monolayer whose compression is accompanied at low coverage by a decrease of the effective interfacial tension, followed at higher densities by interfacial buckling \cite{milner1989buckling, Aveyard2000, monteux2007determining, Pitois2015}. As for a solid beam \cite{landau1975elasticity}, the selected wavelength gives the elastic Young modulus (for 3D approach) or the elastic bending modulus (for a 2D approach) of the particle-laden interface \cite{Vella2004, Planchette2012}.  Yet, in contrast to elastic sheets or molecular assemblies, particle rafts show a strong granular character associated to the development of Jansen effect evidenced both at the raft scale \cite{Cicuta2009} and more locally \cite{saavedra2018progressive}, as well as for not spherical particles  \cite{basavaraj2006packing}.  For bidisperse assemblies, averaging the properties found for small and large particles do not satisfyingly represent neither the bending nor the rupturing behaviors developing when both are employed \cite{pre_bidisperse, planchette2018rupture}. Beside these findings, particle rafts are found  to  plastically fold \cite{jambon_2017} and to fracture \cite{Vella2006}. Fracture strength of polydisperse assemblies cannot be deduced from results obtained with corresponding monodisperse particles suggesting particle packing plays a crucial role \cite{fracture_polydisperse}. Despite significant research on particle raft, the modeling of their mechanical properties remain challenges \cite{protiere_2023} and limit their usage.

This lack of predictive models associated with the irreversible character of interfacial adsorption question the ability of such particles to effectively stabilize interfaces. This is especially critical when high strains at fast rates are at play, i.e. for conditions that are common in numerous industrial processes \cite{Garbin_2019}. Indeed, in contrast to surfactants that can migrate from micelles dispersed in the bulk  to the interface, particles cannot build up such reservoirs  and only those initially found at the interface can contribute to interfacial stabilization. Thus, it has been proposed to store additional particles in the form of folds. Upon sudden increase of interfacial area, the excessive particles may be released and could migrate along the coverage gradient, as for  Marangoni flows, and thus close freshly opened holes in the coverage. This self-healing ability, while poorly investigated, has appeared to be strongly dependant of the compression direction \cite{Plohl_2022}. This dependency indicates the importance of the chain force network in triggering  unjamming \cite{liu1998jamming} and further enabling its progression within the assembly \cite{tordesillas2011}.  It also raises additional questions about the stability and dynamics of this network. First of all, about the structure and geometry of the network: how far does it develop in the direction of the compression and perpendicularly to it? which role plays the raft aspect ratio or raft compression level? are the mechanical properties of the network influenced by the method used to prepare the assembly?  At the level of the individual particles: are all particle-particle contacts comparable in term of stress propagation within the raft? can the strength of such contacts evolve for example via ageing  of the contact line \cite{Kaz_2012, mears_2020} and therefore modify the capillary mediated lateral interactions \cite{Kralchevsky_2001, vella_2005, Botto_2012, xue_2014}? How does this affect raft cohesion and its self-healing ability?

In this paper, we shed light on some of these questions by experimentally studying the relaxation of compressed rafts, and more particularly the effects of raft geometry and raft preparation  on their relaxation behaviour. We especially consider the extend of  unjamming and its consequences for self-healing.

 The method and range of studied parameters are described in the Methods section. The used set-up is similar to the one employed to investigate the effects of compression direction on raft self-healing ability \cite{Plohl_2022}. Care is taken to independently vary the raft length and the opening  width so that information about the network geometry can be deduced. Further,  three types of raft preparation are employed for each configuration to evaluate possible ageing effects on the force chain network or at the particle level via for example changes of the contact line topology. Finally, the raft compression level is varied to identify potential effects on the self-healing dynamics. The methods section ends with a description of performed data analysis.
 
The results, reported in the section Results and discussion, evidence different  types of relaxation, which are first qualitatively described. More precisely, two extreme cases can be identified that correspond either to the quasi absence of relaxation and unjamming, or to their full development. In between, a broad range of intermediate relaxation can be reached, whose magnitude appears to be directly related to the extend of unjamming. Regime maps built on measured unjammed areas and raft compression levels show that relaxation magniture is strongly modulated by both the raft preparation  and the width of the opening at the interface.  Quantitative characterization of raft relaxation is then given based on stress measurements and various  scaling laws of unjammed areas. These results confirm the relation between stress relaxation and unjamming.  The paper then focuses on the transition between these regimes are more particularly on the onset of longitudinal and lateral unjamming. The latters are interpreted  in terms of network stability against shear and elongation. Considerable variations of these thresholds with respect to raft preparation evidence that mixing causes a weakening of the force chain network, in agreement with simulations of jammed granular material \cite{tordesillas2011}. The consequences for self-healing ability are then evaluated with the help of a correlation between the surface getting covered by released particles and  the unjammed area.  The paper ends with the conclusions.

\section{Methods}
\subsection{Experimental setup}

\begin{figure*}
\centering
  \includegraphics[width=0.99\linewidth]{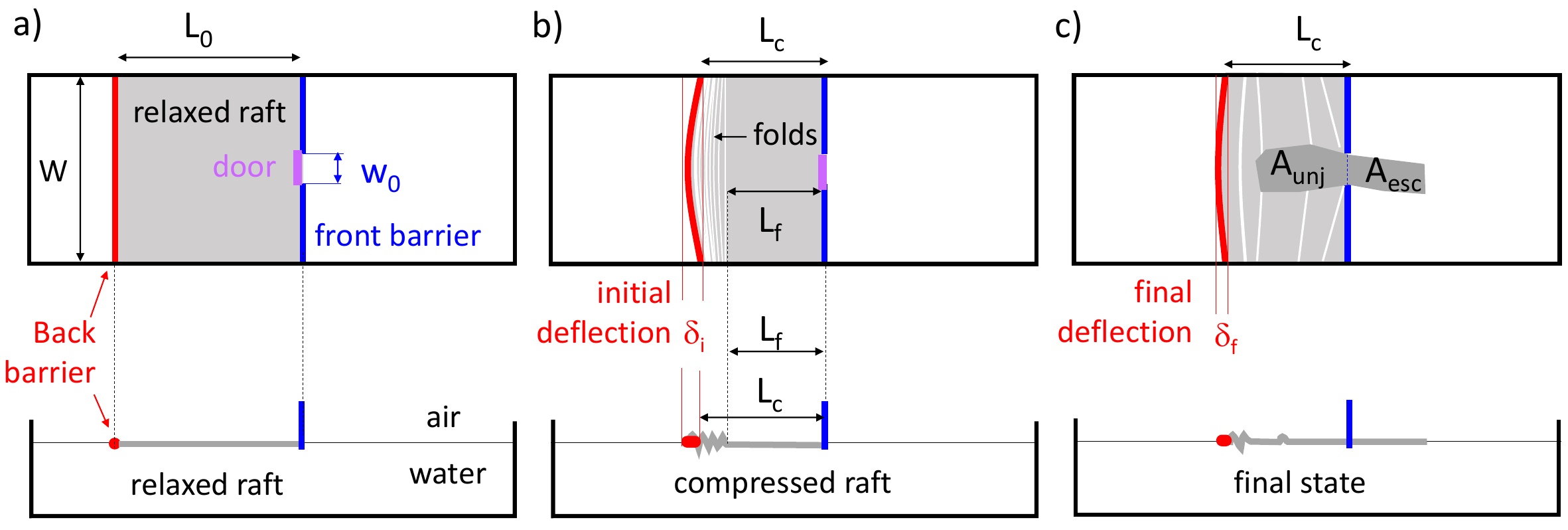}
  \caption{Experimental set-up sketched from  top and side views a) before raft compression, b) after raft compression and before local stress release, and c) after  stress release showing the final raft  state. The through width is fixed and equal to  $W=60 mm$. The raft  is compressed from  the back reducing its length from $L_0$ to $L_c$ using a movable elastic barrier (red) whose deflection  ($\delta_i$) gives access to the initial lineic pressure $\Pi_i$. During this process, folds form at the back, which are separated from the front barrier by a distance  $L_f$. The door blocking the orifice pierced in the front barrier (width $w_0$) is then suddenly opened and the raft evolution followed with a high speed camera placed under the trough. In its final state, the raft unjammed over an area $A_{unj}$ and a surface $A_{escp}$ got covered by particles passing the front barrier (blue).  The final stress $\Pi_f$ is obtained from the final elastic deflection $\delta_f$. Note that the elastic deflection is exaggerated here ($\delta < 0.00023W$).}
  \label{fig_set-up}
\end{figure*}

The experimental set-up - sketched in Figure \ref{fig_set-up} -  consists of a trough to confine capillary adsorbed particles in a controlled manner, as described in \cite{Plohl_2022}. Two fixed lateral walls distant from $W=60mm$ are combined with movable back and front barriers. The back barrier is made of an elastic rubber, whose deflection indicates the local lineic pressure building in the back of the raft. It is translated toward the front barrier to compress  the raft at the desired level. Note that during raft compression, the front barrier remains immobile, in contrast to some of the experiments carried out in \cite{Plohl_2022}. The front barrier is pierced in its center by an orifice of width $w_0$ equal to $4.3$, $9.5$, and $19.9mm$. The door, which initially blocks the orifice can be suddenly opened to locally release the stress at the front of the raft, enabling its relaxation.

The raft relaxation is observed by a high speed camera placed below te through whose bottom plate is transparent.  Typically, the frame rate is set to $3000fps$ and the magnification to $173 \mu m / px$. By analysing the recorded videos (see section \ref{sec:analysis}), we obtain, among others, the area that unjams,  the  rate at which particles escape the initially confined domain, the final area they newly cover, and  the  stress  at the raft back.

\subsection{Preparation protocols}
All the rafts presented in this study are made of the same particles, namely silanized glass beads. Their size distribution is gaussian, centered at $d_{part}=127 \mu m$ with a narrow standard deviation of 5 $\mu m$. The contact angle with distilled water is  $110^{\circ}\pm5^{\circ}$.

\subsubsection{Production of relaxed raft}
These particles are placed in various amounts between the fixed walls and movable barriers and mixed according to three different protocols described below.

The first protocol consists in gently sprinkling the particles at the air/distilled water interface. The space between the barriers is chosen to be sufficiently large for all particles to easily accommodate in the form of a monolayer. To ensure that all particles are effectively adsorbed at  air/water interface, gentle blowing is manually applied  with a Pasteur pipette. At the end of this preparation step, the particle assembly is homogeneous and its packing approaches the random close packing.

In the second protocol, the sprinkled particles are redistributed along the interface with the help of an hydrophobic Pasteur pipette tip. Practically, small particle assemblies that spontaneously form upon lateral capillary attractions \cite{Botto_2012} are carefully brought together until most of the gaps between them get filled. The translation of these assemblies occur in the plane of the interface, i.e. the pipette tip never comes in contact with the bulk water.  This action is source of local shear and compression of the particle monolayer leading to the  formation of small wrinkles that spontaneously relax. The resulting particle assembly  presents almost no holes and its packing is close to  $\pi/(2\sqrt{3})$, the maximum packing obtained for hexagonal crystal. 

The third and last protocol subjects the particles to strong mixing. The adsorbed particles are energetically stirred with a stick whose tip is immersed into the water. The stirring is applied until particle chains  are all broken. The particles then  form  many tiny and quasi monodisperse islets. The resulting assembly shows the presence of holes whose  dimensions are comparable to those of the tiny islets. The latter are in contrast  densely packed.

These three protocols mostly differ from the mixing method applied to the particles, and we refer to the corresponding rafts as to "tempered", "sheared", and "annealed" rafts, respectively. We choose these terms in reference to the facts that particle rafts are known to be soft glasses \cite{prl_protiere_2017}. Thus, the possibility for the athermal particles to rearrange and lower stored internal stress  is directly related to the states they could explore during preparation, and thus to the applied mixing protocol.

Once a particle assembly has been formed, and regardless from the employed protocol,  it is slightly compressed and uncompressed from the back (5 cycles, until apparition/disappearance of the first/last wrinkle). This step is needed to reshape the raft so that it fits in the  square  confined area of the trough.
 The length of the "relaxed" raft, i.e. of the raft observed at limit of back stress detection, is then measured  and refereed to as "relaxed length" noted $L_0$.
 
Note that the amount of particles placed at the interface prior to applying any mixing protocol is weighted and adjusted to produce rafts whose relaxed length  is  $40$, $60$ or $90mm$ with deviations of less than $\pm 1.8mm$ ($<5\%$).

\subsubsection{Raft compression and stress release}

 The relaxed raft is then quasi-statically compressed by slowly translating the back barrier toward the front. The compressed length is then measured and noted $L_c$, see figure \ref{fig_set-up} b). For the short and long rafts, i.e. for $L_0$ equal to $40mm$ and $90mm$, three compression levels are considered, providing values of $K=(L_0-L_c)/L_0$ close to 33\%; 50\% and 66\%. For the initially square rafts ($L_0=60mm$), two additional compression levels are investigated,  a lower one at 25\% and a higher one at 75\%.
 
These compressed rafts are then let to relax by opening the door placed in the front barrier. Orifices with three different widths can be used providing $w_0$ equal to either $4.3$; $9.5$; or $19.9mm$. Each of them is used for every type of raft preparation (tempered, sheared, annealed). Together with the  11 combinations of $L_0$ and $K$, we performed roughly 100 different experiments. 

\subsection{Data analysis}\label{sec:analysis}

The raft, in its  compressed state, is made of a jammed and loacally folded particle monolayer. The folds caused by friction undergone during compression, are found at the back, thus at the opposite side from the front barrier. They appear bright on the pictures, which allows the measurement of $L_f$, the initial distance between the front barrier and the first fold, see figure \ref{fig_set-up} b).

\begin{figure}
\centering
  \includegraphics[width=0.9\linewidth]{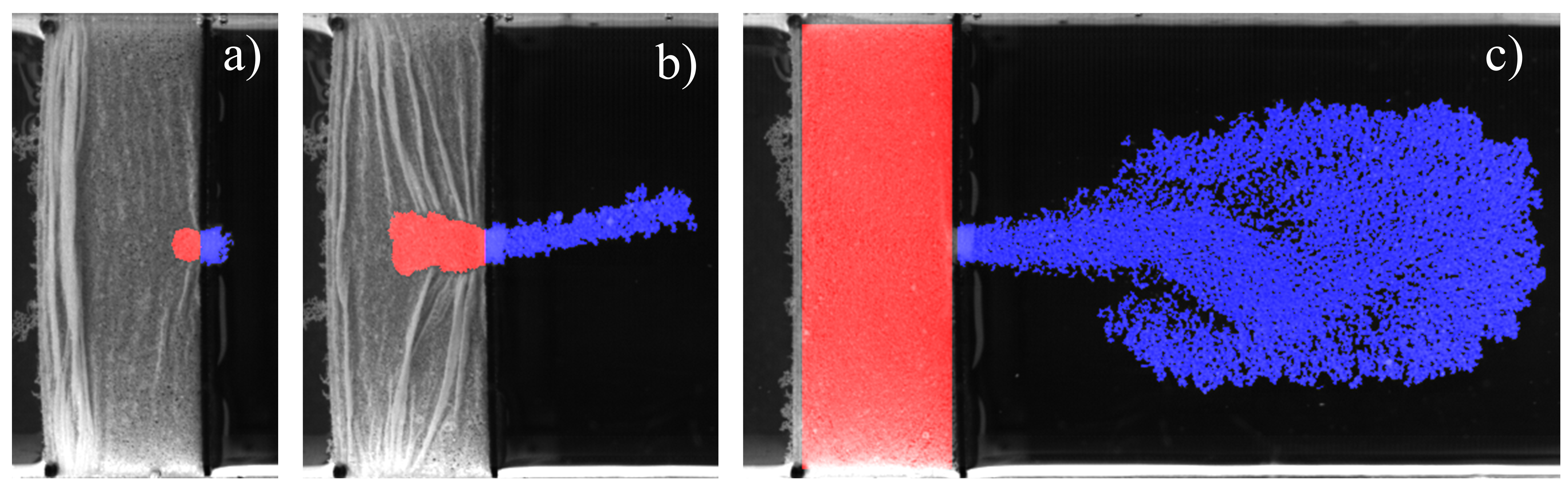}
  \caption{Final raft relaxation for similar geometry ($L_0 \approx 40mm$ and $K\approx 50\%$) and different mixing degrees: \textbf{a)} tempered, \textbf{b)} sheared and \textbf{c)} annealed rafts. Red color denotes $A_{unj}$, the final unjammed area and  blue color shows $A_{esc}$, the final area covered by escaped particles. }
  \label{fig-exp}
\end{figure}

Once the door is opened, the stress locally relaxes, leading to unjamming. The unjammed particles flow through the orifice and reach the initially uncovered interface, where they are considered as "escaped" particles. Depending on the experiment, the unjamming can be almost absent, partial stopping before the back stress significantly reduces, or total leading to the full raft relaxation.
The final unjammed area, $A_{unj}$, is detected automatically using a machine learning pluggin of ImageJ as explained in \cite{Plohl_2022}. The final escaped area, $A_{esc}$, is the dinal surface covered by the escaped particles. It can be automatically detected using a simple threshold function on ImageJ. 
Examples of $A_{unj}$ and $A_{esc}$ detection are given in figure \ref{fig-exp} for three rafts of similar geometries ($L_0 \approx 40mm$ and $K\approx 50\%$) obtained according to different preparation methods. From left to right, $A_{unj}$ (red) and  $A_{esc}$ (blue) are increasing.  

The dynamics of escaping particles is further obtained by analysing high speed videos. The particle flow rate $Q(t)$, is calculated as the temporal derivative of $\Sigma_{esc}(t)$, the surface covered by  escaped particles at the instant $t$, measured from door opening. For a large number of experiments, the flow rate is found to be constant, in agreement with  reported behaviour  of back compressed rafts \cite{Plohl_2022}. Thus, in the following, we will only consider  those constant flow rates, $Q$. The experiments for which a constant flow rate cannot be extracted correspond either to very limited relaxation, or to very fast and total relaxation.

Finally, we also measure $P=\Pi_f/\Pi_i$, the ratio between the final and initial lineic pressure  at the back of the raft, see also figure \ref{fig_set-up}. The relation between the rubber band deflection $\delta$ and the lineic pressure $\Pi$ is recalled bellow :
\begin{equation}
     \Pi=\frac{8\lambda_0}{s_1}
    \left[ 
    \left(1-\frac{s_0}{s_1}\right)\frac{\delta}{s_1}+
    \left(\frac{4 s_0}{3 s_1}-1\right)\left(\frac{2 \delta}{s_1}\right)^3
    \right]
    \label{eq:Force}
 \end{equation}

    where $s_0$ and $s_1$ are  the rubber lengths in absence of stress, and  under the slight pre-stretched condition with which it is fixed to its support. Neither $s_0$ nor $s_1$ are varied during all the presented experiments. The parameter $\lambda_0$ corresponds to the ratio of the rubber Young modulus and its cross section. It remains constant for all measurements and is determined via a calibration process. A detailed derivation of Eq. (\ref{eq:Force}) and of the calibration process can be found in  \cite{Plohl_2022}.

    Thoroughly all carried out experiments, $\Pi_i$, the initial lineic pressure is found to be constant, in agreement with the existence of a buckling (or folding) threshold \cite{Aveyard2000, monteux2007determining, jambon_2017}. The deflection is found to be of $7.5 \pm 0.5$ pixels, thus with variations  in the range of measurement uncertainty.
    
    As the deflection remains small  ($\delta/s_1<2.3\%$), we obtain at first order:
\begin{equation}
     P=\frac{\Pi_f}{\Pi_i}=\frac{\delta_f}{\delta_i}
    \label{eq:force_ratio}
    \end{equation}

\section{Results and discussion}
\subsection{Relaxation degrees}
\subsubsection{Qualitative description}

\begin{figure*}[h]
  \includegraphics[width=0.99\linewidth]{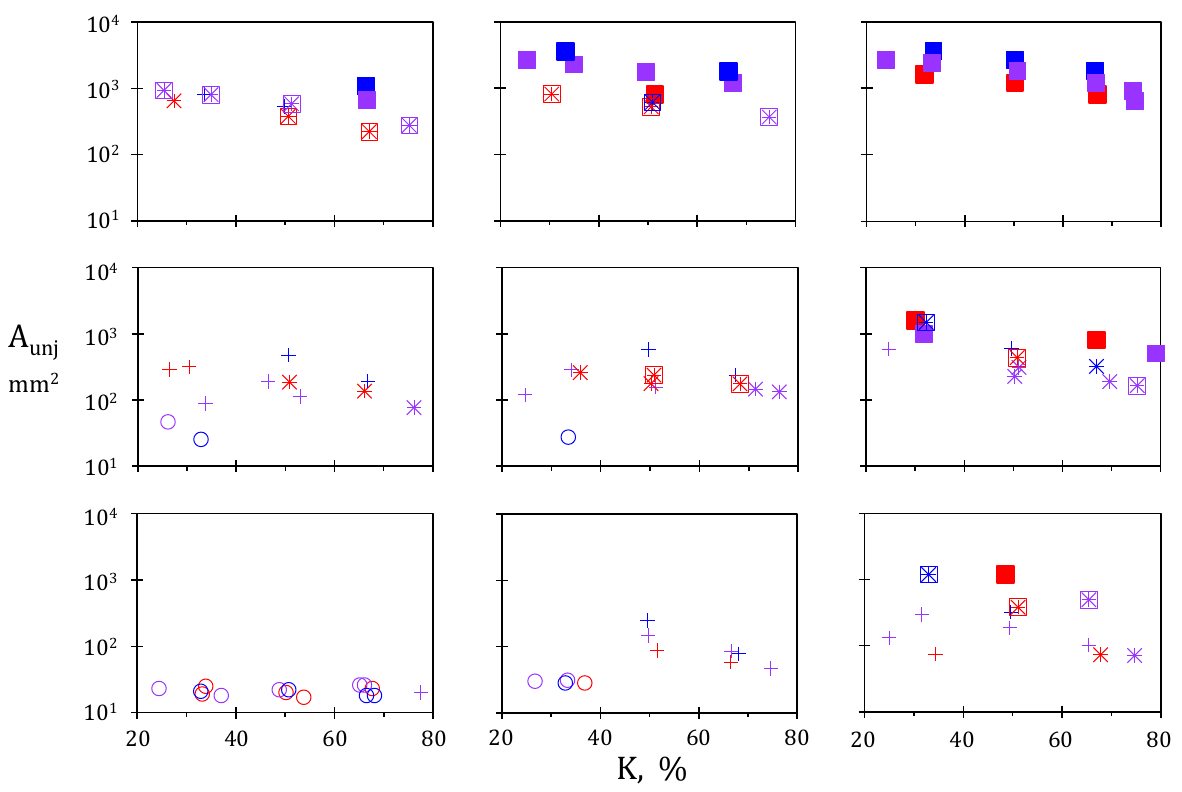}
  \caption{Final unjammed area in $mm^2$ as a function of raft compression for different gate widths and raft preparations. From top to bottom raw, $w_0$ is $19.9$; $9.5$; and $4.3 mm$. From left to right column, the mixing degree increases with tempered, sheared and annealed rafts. Here and in the following colors are used to indicate the relaxed raft length with red for $L_0 \approx 40mm$, purple for $60mm$, and blue for $90mm$. Hollow circles indicate absence of relaxation; simple crosses partial channels; double crosses full channels; double crosses with square  eroded channels; and full squares  total relaxation. }
  \label{fig_regimes}
\end{figure*}    

Compressed rafts experience various degrees of relaxation, which can be associated to different  regimes.

Under the smallest degree of relaxation, only very local unjamming occurs, which remains localised at the gate vicinity, see figure \ref{fig-exp} a). The rest of the raft, especially the folds found at the back, do not show any movement.  No significant variation of the rubber deflection can be detected. Thus, we refer to this regime as \textit{absence of relaxation}. 

For intermediate  relaxation degrees, folds get advected toward the front. The unjammed area extends in the form of  a channel, whose axis aligns with the compression direction. The channel can grow over the entire raft, thus reaching a length equal to the  compressed raft length, or remain smaller, see figure \ref{fig-exp} b).  In the latter case, the back of the raft remains folded or at least jammed on its entire width. In contrast, for channel extending over the entire raft, no fold  remains in the central  section  facing the orifice. The  jammed particles are distributed in two disjoint assemblies that extend on either side of  the channel until the fixed lateral walls. We refer to it as to a \textit{full channel} by opposition to \textit{partial channel}, which stops before the back rubber. 

The next degree of relaxation is reached when further unjamming takes place. Note that  whatever the channel type, the unjamming first extends in the compression direction only. Yet,  after having literally cut the rafts in two, the channel may further grow normally to the compression axis. The removal of the jammed particles resembles an erosion process. Small jammed blocks detach at the channel edge where they instantaneously unjam  and flow toward the gate.
The erosion can  stop before any of the two main jammed blocks vanishes, leaving an \textit{eroded channel}. 

Finally, in certain cases, the erosion causes the jammed blocks found on either side of the channel to totally unjam. Most of the time, the "melting" of the jammed solid-like phase into a "liquid" assembly  occurs for both blocks. Yet, in a few experiments,  only one block melts. For simplicity, we do not distinguish these two situations and refer to both as to \textit{total relaxation}, see also figure \ref{fig-exp} c). 

\subsubsection{Regime maps}
The occurrence of the various degrees of relaxation is represented  in the form of nine regime maps, each of them corresponding to a given combination of raft preparation and gate width, see Figure \ref{fig_regimes}. More precisely, the left column corresponds to tempered rafts, the center one to sheared rafts, and the right one to annealed rafts. Similarly, the lower row corresponds to the smallest gate width, the center one to the intermediate gate width, and the upper one to the largest one. On each map, the final  unjammed area, $A_{unj}$, is  plotted as a function of the raft compression $K$ on \%. The different raft sizes are indicated by different colors: red for $L_0=40 mm$, purple for $60 mm$, and blue for $90 mm$. The degree of relaxation is coded via the usage of different symbols. Empty circles represent \textit{absence of relaxation}; simple and double crosses show \textit{partial }and \textit{full channels}, respectively.  The presence of an empty  square around further indicates \textit{eroded channels} and finally, full squares show \textit{total relaxation}.

The bottom left map representing tempered rafts with the smallest gate $w_0=4.3mm$. Whatever the relaxed raft length and compression level, and except of a single point, only absence of relaxation occurs (hollow circles). On the opposite corner, up right, the results obtained on annealed rafts with the largest gate width $w_0=19.9 mm$ are reported. It is found to always totally relax, as shown by the full squares.

In between, intermediate behaviours are observed, which suggest that both the raft preparation method and the gate width influence its relaxation.
Looking at the effects of the raft preparation, i.e. comparing the various columns, we see from left to right, an increase of the  relaxation degree. For the smallest gate, the evolution goes from almost only absence of relaxation for tempered rafts, to  mostly partial, full, and eroded  channels for the annealed rafts. A single case of total relaxation is visible  for $L=0\approx 40mm$ and $K\approx 50\%$, i.e. for conditions that also led to  eroded channel (repeated experiment), suggesting it should be carefully considered.
For intermediate gate width,  tempered rafts show partial and full channels  coexisting with absence of relaxation. For annealed rafts, the absence of relaxation disappears and several cases of eroded channels and total relaxation can be seen. In the case the largest gate is employed, the absence of relaxation is not found even for tempered rafts. Instead, partial and full channels can be observed, which disappear for sheared rafts already, leading to only total relaxation for annealed rafts.  

Considering the single total relaxation observed with the smallest gate carefully, it appears that the intermediate gate width shows the broadest variety of relaxation degree.

The variety of relaxation degrees can also be seen while comparing experiments performed with different gate widths. Starting with the smallest gate (bottom raw) and going to the largest one (top raw), a clear increase of relaxation degree can be seen. While the bottom raw is dominated by the absence of relaxation and partial channels, the top raw shows mostly total relaxation. The intermediate raw, i.e. the experiments carried out with $w_0=9.5mm$ are dominated by partial, full and eroded channels. Beside a few points corresponding to both  absence of relaxation and total relaxation can be found,  providing the entire variety of relaxation degrees. 

Within the investigated range of mixing degrees and gate widths, the effects produced by  gate width variations, especially the usage of the largest gate $w_0=19.9mm$, appear more pronounced than those caused by changing raft preparation method. The latter are however significant, in particular for the two smaller gate widths. The difficulty to characterize this state, i.e. to quantify the raft mixing degree, makes these effects critical for practical applications. Indeed, our observation suggests that the relaxation and self-healing abilities of rafts are directly related their history, i.e. to a badly controlled parameter. First quantitative evidence and interpretation of this phenomenon are presented below.

\subsubsection {Quantitative regime characterization }

\begin{figure}
  \includegraphics[width=0.5\linewidth]{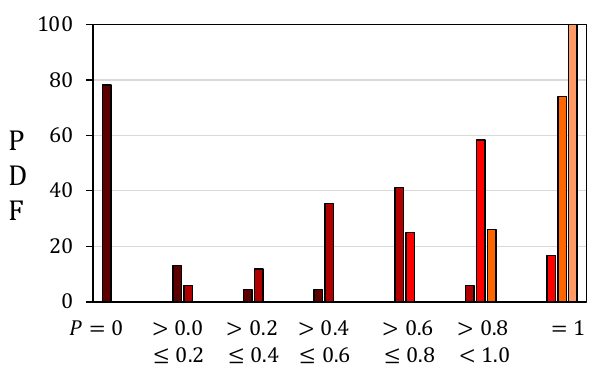}
  \caption{Probability distribution function (PDF) from $P$. From darker to brighter color, the relaxation levels are: total relaxation, eroded channels, full channels, partial channels, absence of relaxation.}
  \label{fig_histo_stress}
\end{figure}

\begin{figure*}[h!]
  \includegraphics[width=0.99\linewidth]{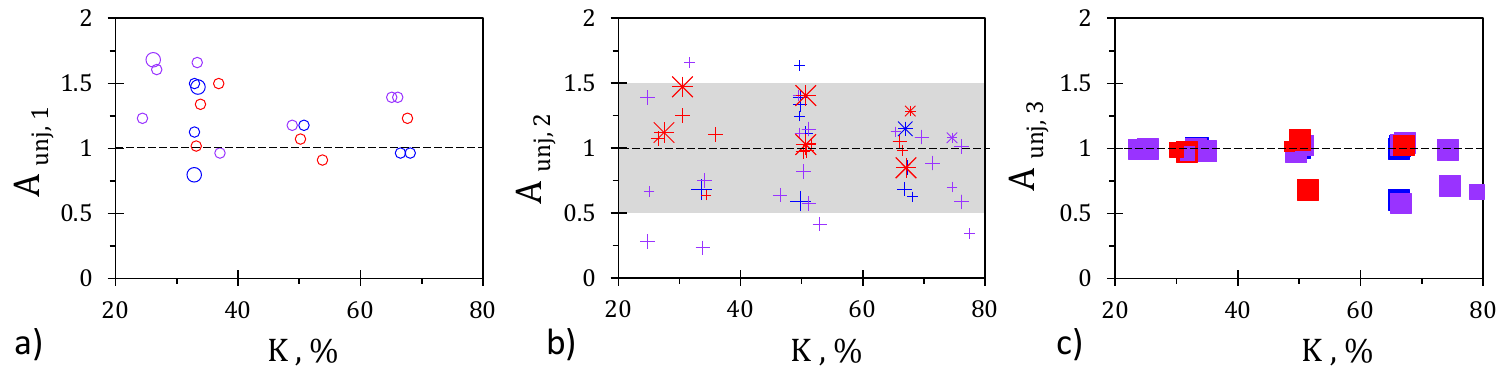}
  \caption{Rescaled unjammed area as a function of raft compression in \%. a) $A_{unj, ~1}$ for absence of relaxation, b) $A_{unj, ~2}$ for partial and full channels, and c) $A_{unj, ~3}$ for total relaxation. For clarity, eroded channels are not plotted. Symbol colors and shapes as in Fig. \ref{fig_regimes}; symbol size indicates  gate width. }
  \label{figure_A_unj}
\end{figure*}

To strengthen our qualitative observations, we plot for each  relaxation regime,  the probability density function (PDF) of $P$, the ratio between final and initial back stress, see Fig. \ref{fig_histo_stress}. The first and last bins correspond, to $P=0.0$ and  $P=1.0$ respectively, while intermediate bins have a width of 0.2. The relaxation degrees are indicated by  color darkness, the lighter standing for absence of relaxation and the darker for total relaxation. Note that each population, i.e. each degree of relaxation, contains at least 12 points providing an acceptable statistical meaning. The PDFs are in agreement with our qualitative interpretation: in  absence of relaxation (24 points), the stress remains always unchanged, i.e. $P=1.0$; partial channels lead to limited stress relaxation with $P<1.0$ for only 26\% of all events (27 points). Significant  stress decrease appears with full channels for which more than 80\% of the 12 measurements show $P<1.0$. This trend becomes even stronger for eroded channels (17 points)  showing always  partially but never totally relaxed stress, i.e. $0<P<1$, with  a broad distribution in between. Finally, total relaxation is associated with almost 80\% of 23 measured rafts for which $P=0$, i.e. of rafts that can truly be considered as totally relaxed. Other points are found below 0.6 and typically correspond to rafts for which only one of the two main jammed blocks "melted". 

According to qualitative regime description, three main scalings  of the final  unjammed area can be proposed. When only local unjamming occurs, i.e. for absence of relaxation, the unjammed area is expected to depend solely of the gate width $w_0$. Corresponding normalization provides :
\begin{equation}
    A_{unj,~1}=\frac{A_{unj}}{{w_0}^2}
    \label{unjam_1}
\end{equation}
The next degree of unjamming is obtained when a channel develops. In this case,  the unjammed area is expected to be proportional to the gate width, $w_0$ and to the compressed raft length, $L_C$. The former roughly fixes the channel width, while the latter gives the maximum length the channel can reach. Thus, this normalization reads:
\begin{equation}
    A_{unj,~2}=\frac{A_{unj}}{{w_0}L_c}
    \label{unjam_2}
\end{equation}
Finally, if erosion proceeds and total relaxation occurs, the whole raft  unjams, leading to:
\begin{equation}
    A_{unj,~3}=\frac{A_{unj}}{W L_c}
    \label{unjam_3}
\end{equation}
where $W$ is the total raft width, fixed in this study to $60mm$.

In Figure \ref{figure_A_unj}, data corresponding to a) absence of relaxation, b) partial or full channels, and c) total relaxation are rescaled with previously introduced scalings and plotted against raft compression $K$. 
As expected, unjammed areas measured for rafts showing no relaxation are satisfyingly rescaled by ${w_0}^2$, see  \ref{figure_A_unj} a). Indeed all values of $A_{unj,~1} $ are in the order of magnitude of  1, which is not the case if employing $A_{unj,~2} $ or $A_{unj,~3}$ (not shown).

For rafts forming partial and full channels, the points are better brought together by a scaling based on  gate width and compressed raft length, see \ref{figure_A_unj} b). As expected, $A_{unj,~2}$ values obtained for full channels are often larger than those measured for partial ones. Despite a certain dispersion,  all points remain close to 1 within  $\pm 50\%$ range (grey domain). While this may look  rough, it is rather good considering the braod range $A_{unj}$ covers (note the logarithmic scale in Fig. \ref{fig_regimes}).  For clarity, eroded channels are not represented. Their rescaled unjammed area, $A_{unj,~2}$, is systematically found  above 1, up to 4 for the narrowest gate, as expected from the qualitative description.

Finally, totally relaxed rafts show an unjammed area that scales  with the total confined area, as proven by  subfigure c) where $A_{unj,~3}$ is employed. A few points deviate from  1, that are systematically found slightly above 0.5. They correspond to  cases where only one of the two main blocks completely erodes.

These observations confirm the existence of various relaxation degrees that are strongly related to the unjamming process. For a fixed raft geometry, i.e. given initial length and compression, relaxation is significantly modified by the gate width and raft preparation method. An interpretation of these results is proposed in the next section.

\subsection{Regime transition}

\subsubsection{Channel formation: network rupture under shear}

\begin{figure*}
  \includegraphics[width=0.99\linewidth]{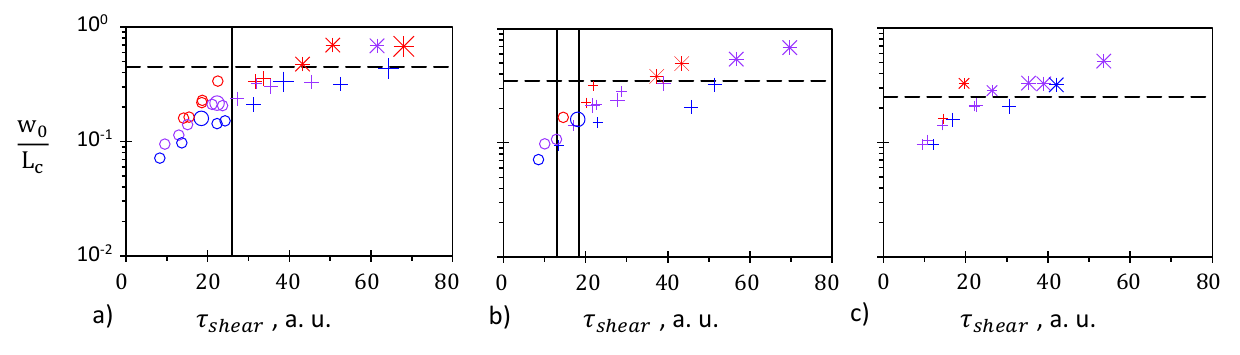}
  \caption{ $w_0/L_c$ as a function of $\tau_{shear}$ for a) tempered, b) sheared, and c) annealed rafts.  Symbol colors, shapes  and sizes as in Fig. \ref{figure_A_unj}. Continuous and dashed lines indicate limits of channel formation and completion, respectively. }
  \label{figure_shear_limit}
\end{figure*}

The transition between absence of relaxation and channel formation can be seen as the rupture of the chain force network, which gets sheared along the compression direction.
To go further, following assumptions must be made.
We consider that, at first order, the stress is fully conveyed by the folds. Beyond, a chain force network develops, which further transmits the back stress in the jammed, yet unfolded, raft part.
We also assume that the distance separating the folds from the front barrier, $L_f$, is only a function of the relaxed and compressed raft lengths. This result is experimentally verified (see Appendix A) and provides the  theoretical length:
\begin{equation}
    L_{f}^{theo}=L_c-0.2051(L_0-L_c)
    \label{l_f}
\end{equation}
The factor 0.2051 is obtained by fitting the experimental data and may vary if different particles are used.
Finally, we  account for the Janssen effect  despite the small raft aspect ratios. In contrast to the work of \citet{saavedra2018progressive} and  in agreement with results obtained in previous studies \cite{Plohl_2022, planchette2018rupture},  we consider that total friction mobilization has been reached. This  can be justified by the application to the rafts of 5 cycles of light compression/decompression   prior performing the desired experiment. In contrast, the results of \citet{saavedra2018progressive} were obtained during first compression, likely explaining the discrepancy. Thus, the force acting on the gate width is expected  to scale as:  
\begin{equation}
    F_{gate} =  w_0 \Pi_i e^{-{L_{f}}^{theo}/\lambda}
    \label{F_gate}
\end{equation}
where ${L_f}^{theo}$ (Eq. \ref{l_f}) is the length on which  the Janssen effect is expected to develop, $\Pi_i$ is the constant initial lineic back pressure caused by  compression, and $\lambda$ is the screening length accounting for friction mobilization on the lateral fixed walls.  The latter has been empirically obtained by fitting measured flow rates, $Q$. It is found to be $42.6 mm$. For more information, please refer to Appendix B.

This force acts in the interfacial plane, whose typical thickness is $d_{part}$. Due to the presence of the front barrier on each side of the orifice, it decreases quickly in this plane on a scale fixed by the force chain network. Thus, $\tau_{shear}$, the shear stress experienced in the orifice vicinity is expected to scale as:
\begin{equation}
    \tau_{shear} = \frac{F_{gate}}{d_{part}\delta} \propto \frac{w_0}{\delta} e^{-{L_{f}}^{theo}/\lambda}
    \label{tau_shear}
\end{equation}
where $\delta $ is a typical length scale, which is expected to vary with the raft preparation method and particle diameter. In our study, the limit between absence of relaxation and partial channel should correspond to a critical value of $\tau_{shear}$, which should itself be a function of the raft preparation method.

Before probing this criterion,   the limit between partial and full channels is considered. Per definition a channel fully develops when its length reaches the compressed raft length,  $L_c$. Moreover, at first order, the channel width is given by the gate width, $w_0$.  Thus, by further assuming that  the length reached by the channel  is  proportional to its width,   $w_0/L_c$ appears a good parameter to assess  the probability for a channel to fully develop. Below a certain critical value, only partial channels should be found, while above, full ones should be observed.   The critical value of this aspect ratio is function of  the chain force network properties, which fixes the coefficient of proportionality between channel length and width. Thus, it may vary with the raft preparation method.

\begin{figure*}[h]
  \includegraphics[width=0.99\linewidth]{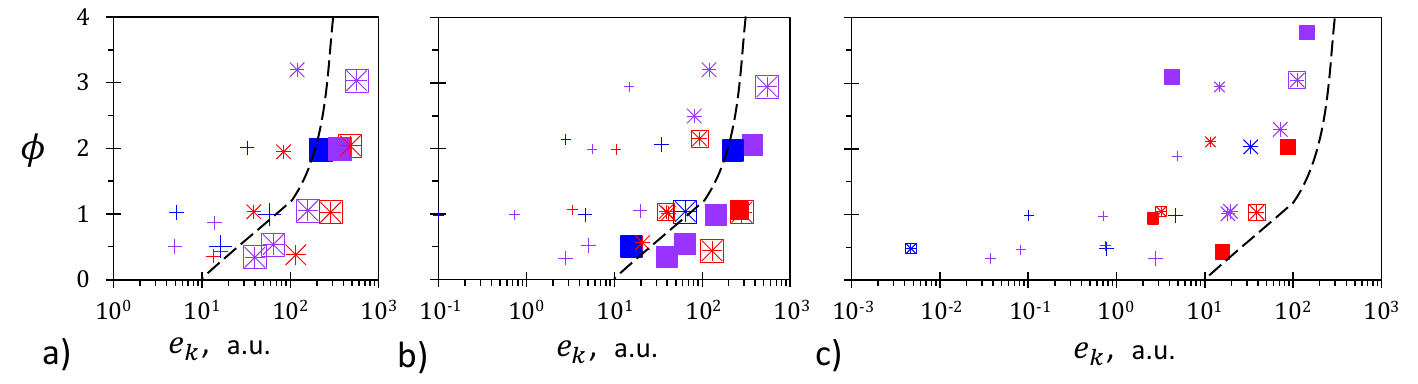}
  \caption{Erosion limit in the form of ($\phi$; $e_k$) diagrams  for a) tempered, b) sheared and c) annealed rafts.  Symbol colors and shapes are as before, symbol size indicates the gate width. The dashed line guides the eye and is unchanged from (a) to (c). }
  \label{figure_erosion}
\end{figure*}

To test this interpretation, we replot the data of Fig. \ref{fig_regimes} corresponding to absence of relaxation (hollow circles), and to partial and full channels (simple and double crosses, respectively) in Fig. \ref{figure_shear_limit}. More precisely,  $w_0/L_c$ is reported  as a function $\tau_{shear}$ for (a) tempered,  (b) sheared, and  (c)  annealed rafts. Symbol shape, color, and size meaning remain unchanged.  
Tempered rafts  show a clear transition between absence of relaxation and partial channels for $\tau_{shear} \approx 26 a. u.$, as marked by the vertical black line. The horizontal dashed line at $w_0/L_c=0.45$ separates the partial and full channels as expected. 

For sheared rafts (subfigure b), the limit of channel formation is less sharp, yet clearly visible with a critical value of $\tau_{shear}$ between 13  and 19 a. u. (vertical lines), thus below the one of tempered rafts. The transition width (6 a. u.) could  be due to the difficulty to reproduce the same level of shear and thus to start with statistically identical networks. The limit between partial and full channels is very well described by a critical value of $w_0/L_c$ of $0.35$, i.e. below the one obtained for tempered rafts. These observations are in agreement with the results of \citet{tordesillas2011}, which show that mixing of granular matter weakens the force chain network and facilitates its unjamming. They are also supported by the hypothesis according to which, at high packing density, unjamming is strongly affected by contact line geometry and mobility \cite{mears_2020}, which are themselves known to be affected by their history \cite{Kaz_2012, xue_2014}.

Finally, annealed rafts (subfigure c) show only partial and full channels. Thus, only channel extension can be checked against our criterion in $w_0/L_c$. The critical value is further reduced, as marked by the dashed line at $0.25$. This decrease further supports the interpretation according to which the more mixed  the rafts are, the less stable the chain force network is, and thus the most probable the formation of a channel extending over the whole raft length becomes.

\subsubsection{Erosion: inertial disintegration of elongated blocks}

Raft erosion looks very different than the progressive unjamming developing along the compression axis. While the latter can be described as a regular and continuous unfolding process, erosion proceeds by successively detaching jammed blocks on either side of the unjammed channel. It is important to keep in mind that after full channel completion, the raft is made of two separated jammed and folded blocks. Indeed, while channel development causes the folds  to move from the back toward the front, it does not cause their disappearance or significantly modify their orientation. Thus, the remaining jammed blocks  present folds whose extremities are located at the channel edges and side walls. Assuming the particle monolayer keeps its elastic properties \cite{Planchette2012a, pre_bidisperse}, the  jammed and folded  blocks are subjected to elongational stress perpendicularly to the channel axis \cite{PetitPHD2014}. The detachment of smaller jammed blocks can thus be interpreted as a failure against elongation. 
The failure is triggered by  inertia of the particles flowing in the channel. The channel inertia  can  be roughly estimated by $E_k \propto L_c w_0 (Q/w_0)^2$. Here $Q$ is the flux of escaping particles, $Q/w_0$, their typical velocity, and $L_c w_0$, the channel surface, which is directly proportional to its mass assuming a constant  particle packing. The erosion occurring for small blocks along the channel edge, we introduce the inertia per unit of channel length:
\begin{equation}
    e_{k} = \frac{E_k}{L_c} \propto \frac{Q^2}{w_0}
    \label{inertia}
\end{equation}

The raft cohesion, which opposes the inertial erosion, can be estimated as the product of (i) particle-particle contact density along the channel edge with (ii) the average strength of particle-particle attraction.  Given the linear compression applied to the raft, the lineic contact density can be estimated by  $L_0/L_c$. Its excess, by reference to the relaxed state, is given by $\phi=L_0/L_c-1$. The strength of the particle-particle attraction is unknown. It is expected to be modulated by the raft preparation.

According to this interpretation, erosion happens above a critical level of $e_{k}$, which increases with increasing $\phi$. To probe the validity of our interpretation, all channels and total relaxation points found in Fig. \ref{fig_regimes} have been replotted as ($e_k$; $\phi$) diagrams  in figure \ref{figure_erosion}. To assess potential effects of  raft preparation, the results are splitted into  three subfigures, corresponding from left to right to tempered, sheared, and annealed rafts. For tempered rafts, erosion is found above critical $e_k$ values, which increase with $\phi$ (dashed line to guide the eyes). The evolution is not linear, yet it is in agreement with a rupture of particle-particle contact by excessive inertia . The dashed guide line drawn for tempered rafts is reproduced in the other subfigures to facilitate comparison. For sheared rafts, the limit of erosion  seems to be shifted toward lower values of $e_k$ for given $\phi$. Subfigure c) obtained for  annealed rafts, shows even more erosion events on the left side of the dashed line. Yet, the limit is not  sharp anymore. No systematic trend can be seen to attribute the dispersion to raft length or gate width. Instead, we suspect the preparation protocol or its effects not to be well reproducible. Beside the limited number of points and a certain dispersion in c), which do not allow for a definitive conclusion, our interpretation appears to be satisfying. Thus, raft mixing seems  not only to lower  resistance against shear but also  against elongation.

\subsection{Consequences for escaped particles}

\begin{figure}[h]
\centering
  \includegraphics[width=0.99\linewidth]{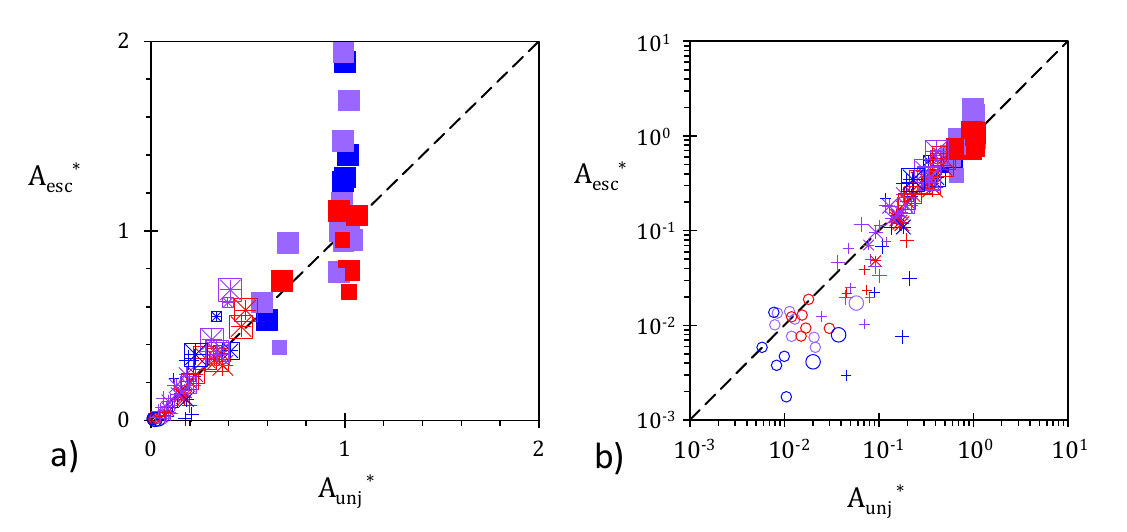}
  \caption{Rescaled area covered by escaped particles, ${A_{esc}}^*$ as a function of rescaled unjammed area ${A_{unj}}^*$.  a) linear scales and b) log scales. Symbol colors and shapes are as before, symbol size indicates the gate width.  }
  \label{figure_esc}
\end{figure}

Capillary adsorbed particles cannot be stored in the bulk in contrast to surfactants, which form micelles above the critical micellar concentration. Thus, while particles can very well stabilize interfaces of fixed or decreasing areas \cite{monteux2007determining, Abkarian_2013}, their stabilizing ability becomes rather limited for extending interfaces \cite{planchette2012transition, planchette2018rupture}. This aspect, which can be advantageously used for certain applications such as drug delivery, remains a bottleneck for other processes \cite{Garbin_2019}. By folding the particle monolayer, particle reservoirs can be obtained from which excessive particles could be released to cover  opening holes \cite{Plohl_2022}. The efficiency of this strategy is related to the amount of released particles and to their migration dynamics. We focus here on the first aspect and evaluate it by plotting ${A_{esc}}^*$,  the normalized area covered by escaped particles  as a function of ${A_{unj}}^*$, the normalized unjammed area, see figure \ref{figure_esc}. The normalization of the unjammed area is made with respect to the compressed raft area, i.e. to $W L_c$, providing ${A_{esc}}^*={A_{esc}}/(W L_c)$. In contrast, the escaped area is compared to its maximum value given by $W (L_0-L_c)$. Thus we define ${A_{unj}}^*=A_{unj}/(W (L_0-L_c)) $.
A good correlation between these two normalized quantities can be seen. Consequently, a first reasonable estimation of the surface getting covered by escaped particles is obtained, which reads:
\begin{equation}
    A_{esc} = \frac{L_0-L_c}{L_c} A_{unj} 
    \label{esc}
\end{equation}

Discrepancies are mostly observed for  small values of ${A_{unj}}^*$ and large values of  ${A_{esc}}^*$. The former can be well explained by the  experimental uncertainty on ${A_{unj}}$  for rafts which do not relax, i.e. where this area remains small and not well visible. The latter can be explained by a change of particle coverage after escape. On one hand, for rafts that fully relax, and more generally for some rafts that are subjected to erosion, the escaped particles do not form a dense assembly. With the used magnification (approx 1.5 particle per pixel), the  thresholding method employed to detect the  escaped particles include  voids between them, which should be excluded. On the other hand, per definition, the unjammed area cannot exceeds the confined area used for normalization and ${A_{unj}}^*$ cannot exceed 1.

Further accounting for the three scalings found for $A_{unj}$, see Eqs. (\ref{unjam_1}), (\ref{unjam_2}), and (\ref{unjam_3}), and for the limits between these three regimes, see  Figs. \ref{figure_shear_limit} and \ref{figure_erosion}, a first reasonable prediction of raft self-healing ability has been obtained in this work. Many open questions remain, which should be addressed in further investigations. One of the most critical is  to understand how raft mixing facilitates network ruptures and thus raft unjamming. Are these effects caused by changes in the network topology such as shorter chains, less branching,.. or by the modification of particle-particle interactions? The latter being mediated by the deformed liquid interface \cite{Kralchevsky_2001, liu2018capillary}, it is probable that raft history  influences contact line  ageing \cite{Kaz_2012, xue_2014}. Different particles of various shapes and surface properties could be used to investigate this point. Ideally, the evolution of the contact line should be measured at the sub-micrometric scale, which remains a challenging task \cite{basavaraj2006packing, loudet2006wetting}.

\section{Conclusions}
Relaxation of back compressed rafts through orifices located at the front side have been studied systematically. While the raft width ($W=60mm)$ and  particle size ($d_{part}=127\mu m$) remain unchanged, raft length, compression level, gate width and raft preparation methods have been varied in the following way: $40\leq L_0 \leq90mm$, $25\leq K \leq 75\%$, $4.3 \leq w_0 \leq 19.9mm$, with  preparation of tempered, sheared, and annealed rafts obtained by applying no mixing, gentle interfacial shearing, and vigorous stirring, respectively.   
Different degrees of relaxation have been observed ranging from the absence of relaxation, mostly found for small gate width and tempered rafts, to total relaxation, mostly occurring for large gate width and annealed rafts. In between, intermediate relaxation develops whose magnitude is strongly related to unjamming extend. The latter is modulated by the gate width but also by the raft preparation. The unjamming process follows a sequential evolution: first development of a channel of width the gate width. The  channel extends along the compression axis and may reach (or not) the back of the raft. Then, if further unjamming occurs, it happens according to a lateral erosion process. Small jammed pieces detach from the two main blocks into the channel where they instantaneously unjam. This evolution  may stop before being completed, leaving an eroded channel and a partial stress relaxation, or continue until  their complete disappearance, which is associated to  total stress relaxation. The limit of channel formation is interpreted as  force chain network rupture against shear. The one of erosion is seen as its rupture against elongation, which is triggered by  escaping particles  inertia. Both limits are affected by the raft preparation indicating that mixing reduces raft robustness and therefore increases its self healing ability. The physical processes causing these effects may be attributed to the modification of the network topology or to modifications of particle-particle capillary lateral interactions via, for example, ageing of the contact lines. Further investigations are needed to better understand the microscopic origin of these effects.

\section*{Appendix A: Distance between front barrier and folds}\label{app_A}

\begin{figure}[b]
\centering
  \includegraphics[width=0.5\linewidth]{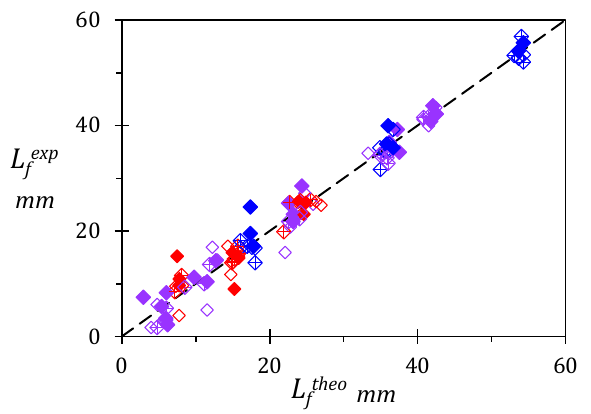}
  \caption{Measured $L_f$ values as a function of ${L_f}^{theo}=L_c -\alpha (L_0-L_c)$ for $\alpha=0.2051$. Colors indicate the relaxed raft length with red for $L_c\approx 40mm$, purple for $L_c\approx 60mm$, and blue for $L_c\approx 90mm$. The symbols indicate raft preparation with empty diamonds for tempered, crossed diamonds  for sheared, and full diamonds for annealed rafts.}
  \label{figure_lftheo}
\end{figure}

During raft compression, folds form at the back barrier, which can be easily distinguished on the recorded images. As Janssen effect is expected to develop in the not folded section of the monolayer, it is important to measure and model the length of the not folded domain. The latter corresponds to $L_f$, the distance separating the folds from the front barrier. The only relevant length scales are the relaxed and compressed raft lengths. We further expect the length of the folded domain to scale with the particle excess, i.e. with $L_0-L_c$. Thus,  a model in the form of: 
\begin{equation}
{L_f}^{theo} = L_c - \alpha (L_0-L_c)
\end{equation}
is considered. Using the least mean squares fitting function of Matlab, we obtain $\alpha=0.2051$. The good agreement between the theoretical prediction and the measured values can be seen in Figure \ref{figure_lftheo}, where all points have been reported, independently from employed gate width or raft preparation method.

\section*{Appendix B: Flow rate modeling}\label{app_B}

For most experiments, the surface covered by escaped particles increases linearly with time indicating a constant flow rate.  Note that experiments with no constant flow rates mostly correspond to absence of relaxation or to total relaxation. Experiments for which  $Q$ is found to be constant are used to search for a possible model. The latter does not aim to be universal but mostly to cover intermediate relaxation cases for which channels form. Since usage of the sand-hour, granular flows are known for their steady character \cite{Brouwers_2007}. In  case Janssen effect takes place, and if the flow is caused by gravity,  the Beverloo law has proven its  validity \cite{Beverloo_1961}. Yet, in our experiments, the flow is triggered by the back lineic pressure, which is expected to be partially transmitted along the compression direction, producing a front lineic pressure at the orifice. As in \cite{wang2021silo, Zheng_2021}, we assume the flow rate to scale linearly with the  pressure at the opening. Force chain developments and friction mobilization at the walls suggest that the front pressure, i.e. the pressure perceived at the orifice, scales as $\Pi_i exp(-{L_f}^{theo}/\lambda)$. $\Pi_i$, the initial back pressure is maintained as long as folds exist. $\lambda$, the length-scale accounting for its progressive screening is a priori unknown and must be deduced from experimental measurements.  
To go further, another striking property of granular flows  must be considered. It is their independence toward material properties. While being discussed by \citet{Madrid_2018}, this assumption is taken for our model. Said differently, despite suspecting that raft preparation modifies the particle-particle interactions and force chain network, no distinction is made between  differently prepared rafts. However, flow rate variations with the orifice width and/or particle size are expected. The most common way to account for them has been proposed by Beverloo already and reads as $(w_0-k d_{part})^{D-1/2}$ with $k$ a numerical coefficient to be empirically adjusted  and $D$ the problem dimension, 2 in this paper. For large orifice widths,  $k$ can be taken equal to 0. More complex variations have also been proposed \cite{ janda2012flow, hu2019size}. All these expressions have in common to keep the dependency with the gate width in  the form of $w_0^{D-1/2}$, which we  assume to be valid in our model. Yet, fitting experimental flow rates with $A {(w_0 -k d_{part})}^{(3/2)}exp(-{L_f}^{theo}/\lambda)$ shows that - whatever the value of $k$ - the variations with $w_0$ are not properly reproduced. Thus, we decide to fix $k$ to zero and to account for orifice width effects via an exponential contribution in $exp(-{L_f}^{theo}/w_0)$. As the shear stress  in the channel is expected to be proportional to the velocity gradient, the scaling in ${L_f}^{theo}/w_0$, can be a posteriori interpreted  as a  frictional or viscous contribution. Note that we do not aim for deeper interpretation here. The modeling goal is mostly to interpret transitions between various relaxation regimes and thus to reproduce the experimental data.

Altogether, we obtain:
\begin{equation}
    Q^{theo}= A~{w_0}^{1.5}~e^{-{L_f}^{theo}/\lambda}~e^{-B~{L_f}^{theo}/w_0}
    \label{Q}
\end{equation}

\begin{figure}[h]
\centering
  \includegraphics[width=0.7\linewidth]{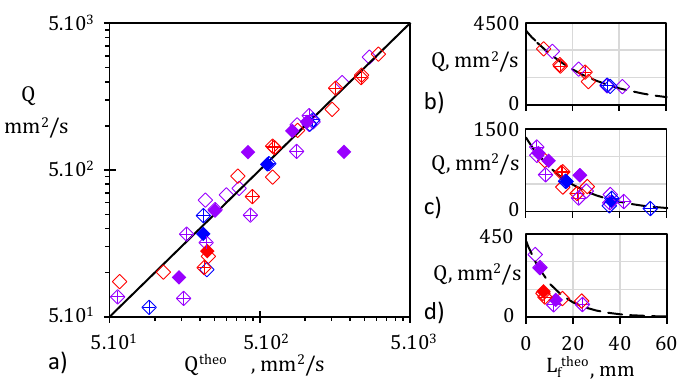}
  \caption{a) Experimental $Q$ values (in $mm^2/s$) as a function of the prediction given by Eq. (\ref{Q}). b), c), and d) Variations of $Q$ (in $mm^2/s$) with ${L_f}^{theo} $ for $w_0$ equal to $19.9$, $9.5$ and $4.3mm$, respectively. Red, purple, and blue colors for $L_0$ of  $40$, $60$ and $90mm$. Empty, crossed, and full diamonds for tempered, sheared, and annealed rafts, respectively.}
  \label{figure_Qtheo}
\end{figure}

A fitting procedure run on Matlab provides $A=3.328$, $\lambda=42.6mm$, and $B=0.2539$. The agreement between the experimental values of $Q$ and those of $Q^{theo}$ predicted by Eq. (\ref{Q}) is very good (correlation coefficient of 0.973), see Figure \ref{figure_Qtheo} a). The subfigures b-d show $Q$ as a function of ${L_f}^{theo}$ for decreasing gate width (top $w_0=19.9mm$, center $9.5$ and bottom $4.3$). Results obtained with each gate width are equally good modeled, indicating that the factor $exp(-B~{L_f}^{theo}/w_0)$ is appropriate here. Note also that is therm is not considered while evaluating the force at the opening. This can be easily justified by the fact that frictional effects come to play when flow is established but do not modify the rupturing threshold.  The colors indicate the raft length  and the symbols the preparation method with the same convention as in Fig. \ref{figure_lftheo}. No systematic deviations for any of these characteristics can be seen.

\section*{Author Contributions}
C.P. conceptualization, funding acquisition, methodology, formal analysis, and writing (original draft, review, editing); G.P. investigation, data curation, formal analysis, and writing (review and editing)

\section*{Conflicts of interest}
``There are no conflicts to declare''.

\section*{Acknowledgements}
We would like to thank the Austrian Science Fund (FWF)
for financial support under Grant No. P33514-N. We also
acknowledge Graz University of Technology for its technical
support, especially the Institute of Hydraulic Engineering and
Water Resources Management for lending us the high-speed
camera and the Institute of Process and Particle Engineering
for helping us to sieve the particles.

\bibliographystyle{unsrtnat}
\bibliography{particles}  






\end{document}